\documentclass[12pt,preprint]{aastex}

\slugcomment{Resubmitted to AJ.}

\shorttitle{Galaxies in front of Quasars}
\shortauthors{K\"onig et al.}

\begin{document}


\title{Galaxies in front of Quasars: \\
Mrk~1456 and SDSS J114719.90+522923.2}


\author{Brigitte K\"onig, Regina
         E. Schulte-Ladbeck and Brian
         Cherinka} \affil{Department of Physics and Astronomy,
         University of Pittsburgh, 3941 O'Hara St.,
         Pittsburgh, PA 15260}

%
%

\begin{abstract}
The chance projection of the disk of Mrk~1456 onto a background QSO is
similar to the case of SBS~1543+593/HS~1543+5921. Mrk~1456 is a
luminous, late-type spiral at $z \approx 0.05$. Though the QSO, SDSS
J114719.90+522923.2 at $z \approx 2$, has not yet been observed with
ultraviolet spectroscopy, it shows strong Ca~II absorption at the
redshift of Mrk~1456 which gives evidence that it is a possible Damped
Lyman Alpha absorber. Spectroscopy of the star-forming nucleus of
Mrk~1456 allows us to apply emission-line diagnostics to infer the
chemical abundances at the center of the galaxy, and to make a
prediction of the expected metallicity on the sightline to the QSO.
\end{abstract}





\section{Introduction}

The processes by which gas cools and collects into dark matter halos,
to form stars and luminous galaxies, are topics of great interest in
galaxy evolution studies. High column density gas clouds, traced
through the Damped Lyman Alpha (DLA) absorption lines seen in the
spectra of QSOs \citep[see][]{wolfe2005}, appear to be the main
reservoir of cool, neutral gas in the Universe.  They are also
believed to evolve and form into typical, present-day, disk galaxies
\citep{wolfe1986, storrie1996, kauffmann1996, mo1998, storrie2000,
cen2003}.

DLAs have enabled us to follow the evolution of the bulk of the
neutral gas in the Universe regardless of whether it emits light or
not. For redshifts greater than about 1.6, the Lyman alpha line shifts
into the optical, where, together with a range of diagnostic metal
absorption lines, it can be observed with ground-based
telescopes. DLAs have afforded us with information on the
line-of-sight kinematic properties of the gas \citep{briggs1985,
prochaska1997}, its element abundances, \citep{lu1996, pettini1997,
prochaska2003d}, and star-formation rates \citep{wolfe2003}. The
derived gas properties are, in some respects, similar to those found
in giant spirals in the local Universe, while in others, they resemble
more closely those of local dwarfs. Unfortunately, we know almost
nothing about the galaxies in which this gas resides. The column
density of $N_{\rm HI} > 10^{20.3}$ atoms cm$^{-2}$, that defines a
DLA system \citep{wolfe1986}, has physical significance because it is
typical of the threshold column density for the onset of star
formation \citep{kennicutt1998}. Accordingly, DLAs should be
associated with visible galaxies, the so-called DLA galaxies. Yet the
detection of DLA galaxies has proven difficult at all redshifts, and
the host-galaxy nature of the DLAs and their evolution with redshift
remains an open question \citep{haehnelt1998, cen2003, nagamine2004a,
nagamine2004b}.

At high redshifts, the detection of a DLA galaxy is difficult because
of the brightness of the QSO and the small angular separation between
the QSO and the galaxy \citep[c.f.][]{wolfe2005}. While over 600 DLAs
with $z>1.6$ are available for follow-up observation
\citep{wolfe2005}, only eight DLA galaxies have been detected in
emission \citep{pettini1995, moller2002, weatherley2005}. The DLA
galaxies exhibit a range of regular and irregular morphologies, which
is perhaps not a surprise considering that the galaxies' images are in
the rest-frame ultraviolet \citep{moller2002}. Similar morphologies
are seen among the Lyman Break Galaxies (LBG) \citep{moller2002}, a
class of high-redshift galaxies selected by their emission
properties \citep{steidel1992}. At present, the relationship
between the absorption-selected DLA galaxies and the emission-selected
LBGs remains to be determined \citep{colbert2002, moller2002}.

While DLA galaxies should be more easily detected at low redshifts,
follow-up searches for the visible counterparts of $z<1.6$ DLAs have
not been entirely successful either. There are many fewer DLAs
available to study at lower redshifts, because the interception
probability per unit redshift at low-z is reduced, and because the
Lyman alpha line can only be observed using the scarce resource of
space telescopes \citep{wolfe2005}. Of the 24 DLA systems known at
redshifts below 1.6, only about half have had the corresponding DLA
galaxy identified.  Less than half of those turn out to be luminous
spiral galaxies with the remaining detections (as well as the
non-detections) being interpreted as low surface brightness galaxies
or dwarf galaxies \citep{lebrun1997, chen2003, rao2003}. It should be
noted that a potential bias against finding luminous spiral galaxies
among low-z DLAs may have arisen from the very fact that DLAs have
traditionally been selected by targeting bright QSOs
\citep{wolfe2005}, which, historically, have been identified based on
their stellar-like appearance on sky survey plates
\citep[e.g.,][]{schmidt1983, hazard1991, hewitt1993, veron2003}.

An obvious approach to uncover the connection between the DLAs and the
galaxies in which they reside, is to search for DLA lines that {\it known}
low redshift galaxies cause in the spectra of more distant QSOs. This
is not a new idea \citep[see e.g.,][]{burbidge1971, bowen1991}, but an
idea that we revived given the vast numbers of low-redshift galaxies
and of QSOs found in the Sloan Digital Sky Survey
\citep[SDSS,][]{york2000}. This methodology to search for DLAs based on
a visible galaxy clearly has its own biases. Specifically, since the
SDSS main galaxy sample is magnitude limited \citep{strauss2002},
our selection favors the detection of giant over dwarf DLA galaxies
and in that respect, may be considered complementary to the
traditional selection method. To this end, we correlated the SDSS DR4
QSO catalog \citep{schneider2005} with the DR4 main galaxy catalog
\citep{strauss2002} and inspected the resulting galaxy-QSO
matches. We reduced this list by considering only those galaxy-QSO
pairs for which (a) the impact parameters are smaller than the
galaxy's Petrosian radius, (b) SDSS spectra are available for both
the galaxy and the QSO, (c) $z$(QSO)$>z$(galaxy), (d) the QSO's spectrum
exhibits a strong CaII doublet at the redshift of the foreground
galaxy \citep{cherinka2007}. Thus we identified the luminous spiral
galaxy Mrk~1456 and its background QSO.

Mrk~1456 (=SBS1144+527A=SDSS~J114720.20+522918.6, $z=0.0476$) is
located in front of the QSO SDSS~J114719.90+522923.2 ($z=1.990$). The
QSO sightline intercepts the galaxy within its stellar disk. This
chance projection of a background QSO with a galaxy at low redshift is
reminiscent of the galaxy-QSO pair SBS~1543+593/HS~1543+5921
discovered by \cite{reimers1998}. SBS~1543+593 is the only DLA galaxy
which has been studied in detail in emission as well as in absorption
\citep{bowen2005, sl2004, sl2005}. Before \cite{reimers1998}
pronounced the object a galaxy-QSO pair, SBS~1543+593 was thought to
be a Seyfert galaxy with the QSO constituting its active nucleus. In
an interesting parallel, \cite{takase} list Mrk~1456 as an
ultraviolet-excess galaxy with a ``double nucleus". SBS~1543+593
turned out to be a DLA when Hubble Space Telescope (HST) was trained
on HS~1543+5921 for follow-up ultraviolet spectroscopy
\citep{bowen2001}.

Here, we present an analysis of photometry and spectroscopy of
Mrk~1456, and show that it is a giant spiral with an oxygen abundance
that is typical of its luminosity. We also examine the QSO's spectrum
for absorption lines from the foreground galaxy, and, based on the
strength of the CaII~K line, find that Mrk~1456 is a candidate DLA
system. We discuss our findings in the context of chemical abundances
for star-forming galaxies (SFG) and DLAs, and make suggestions for
future observations.




\section{The Data}
\begin{figure}
\includegraphics[angle=0,scale=0.85]{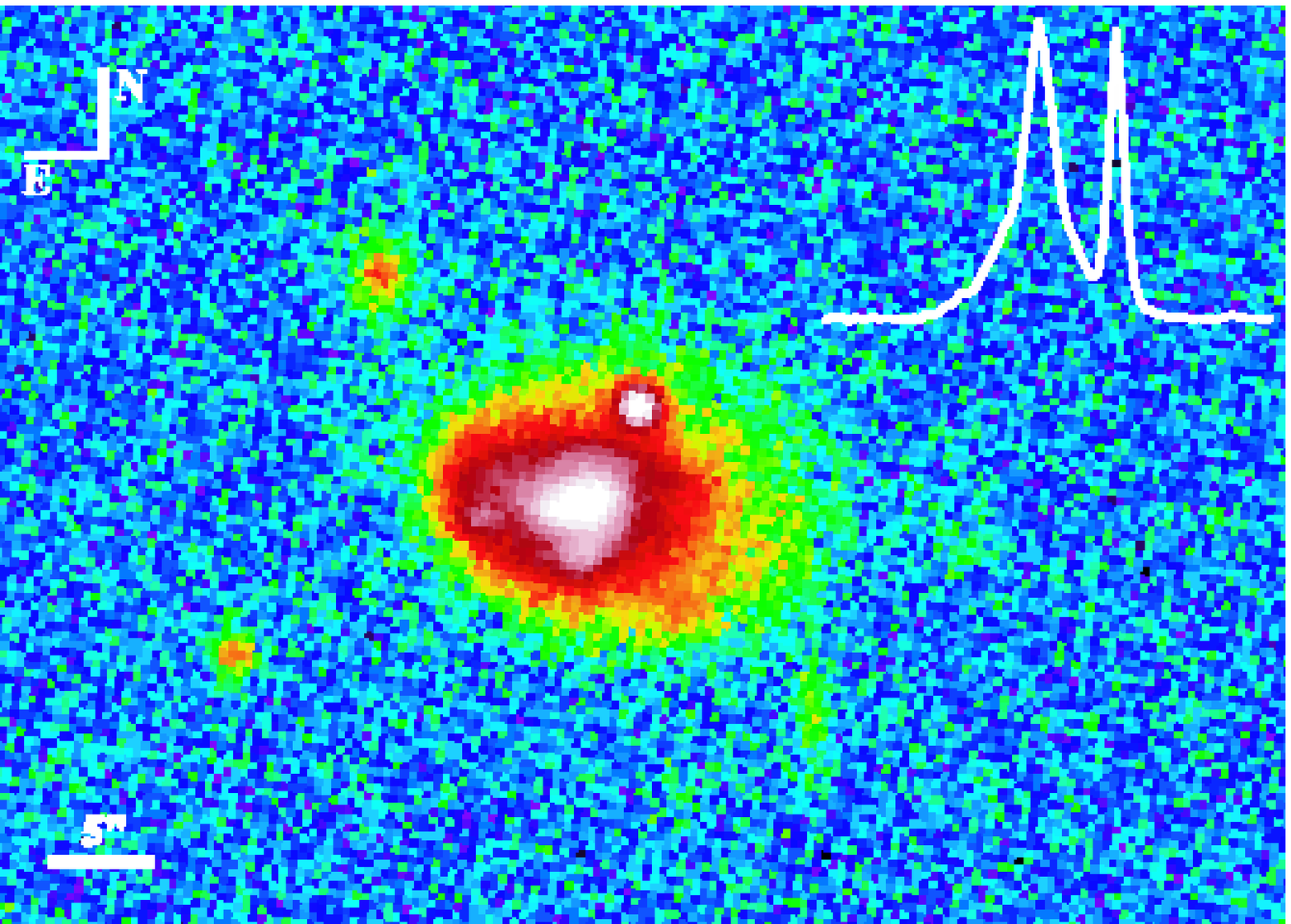}
\caption{SDSS r-band image of Mrk~1456 created using fits files from
the SDSS homepage. The fibers for the spectroscopy are centered on the
nucleus of the galaxy and on the QSO. The inserted white line (right
upper corner) shows a trace of the image through the center of
Mrk~1456 and the QSO.\label{fig:image}} 
\end{figure}

\begin{figure}
\includegraphics[angle=90,scale=.55]{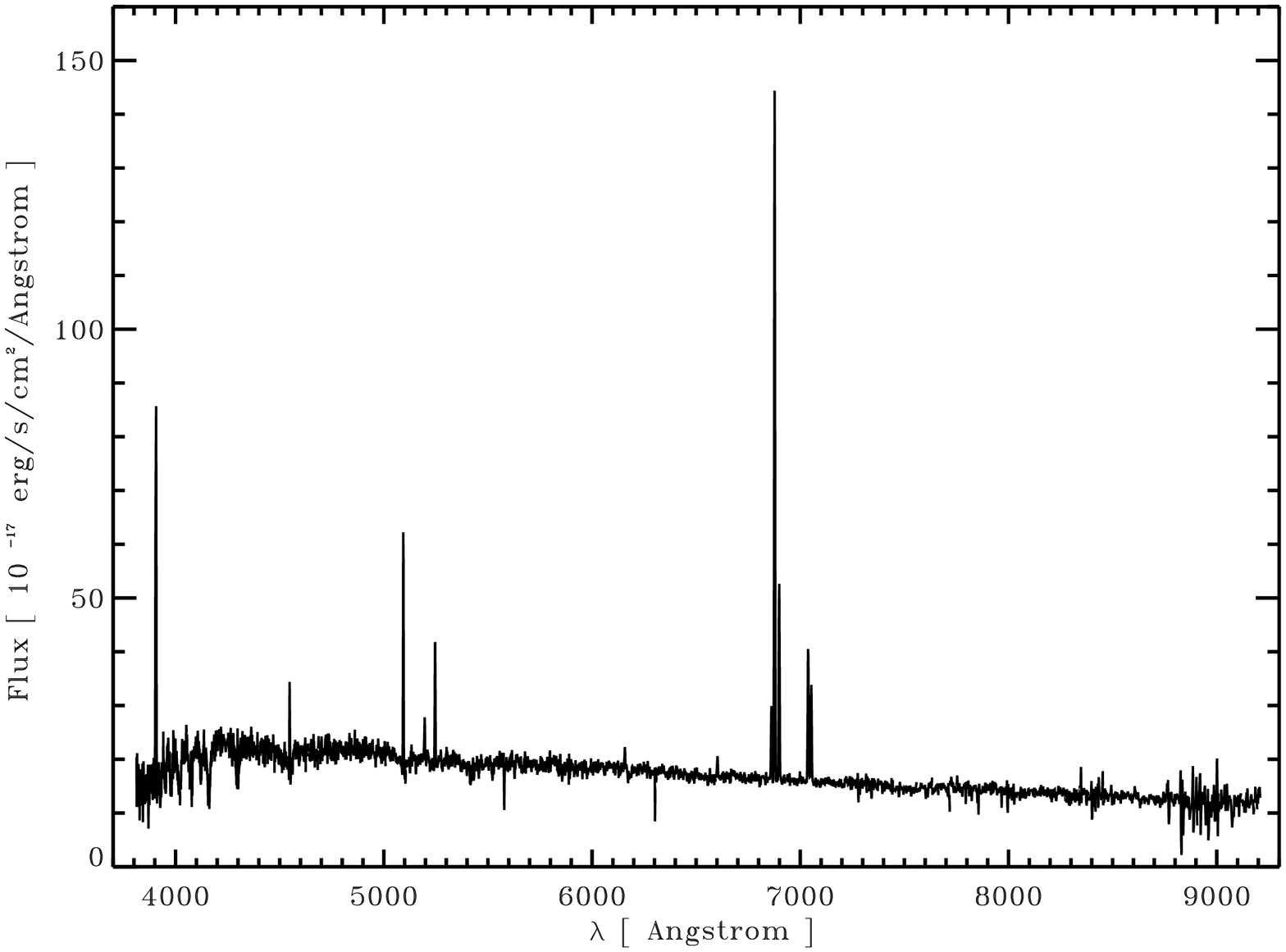}
\includegraphics[angle=90,scale=.55]{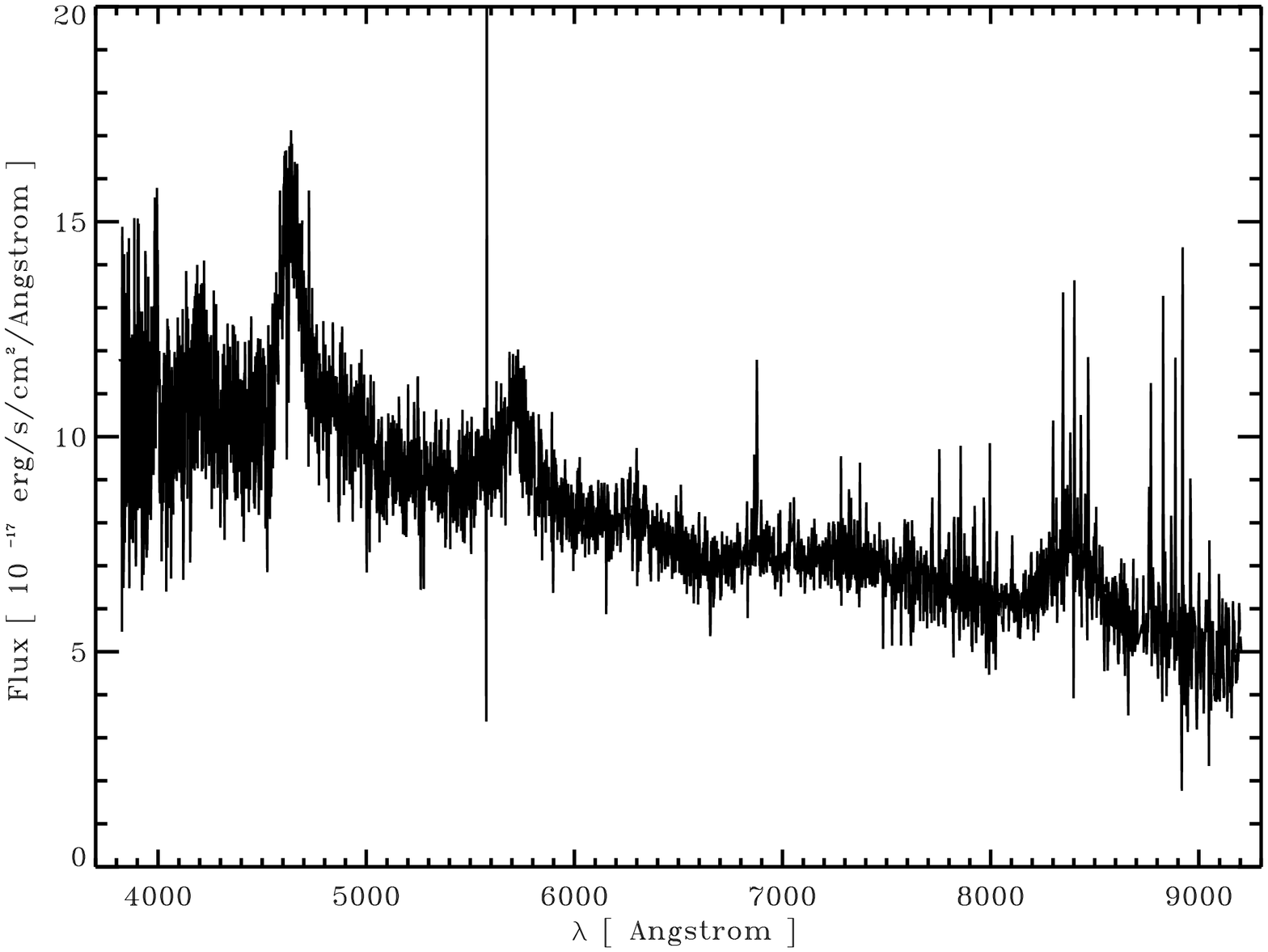}
\caption{The upper plot shows the SDSS spectrum of
  Mrk~1456 \label{fig:spec_g} and the lower plot the spectrum of the
  QSO SDSS~J114719.90+522923.2 from the SDSS.}\label{fig:specall}
\end{figure}

For both, the QSO and the galaxy, we use the pipeline reduced and
flux-calibrated data from the SDSS. Photometric and spectroscopic data
are available for both objects. The fibers were placed on the center
of the galaxy and on the QSO (see also Fig.~\ref{fig:image}). The QSO
sightline intercepts the disk of the galaxy at a separation of
5.3\,arcsec from its center or at a physical distance of 4.9\,kpc
($H_0 = 70\,{\rm km}\,{\rm s}^{-1} {\rm Mpc}^{-1}$, $\Omega_m = 0.3$,
and $\Omega_{\Lambda} = 0.7$). The QSO is in the background at a
redshift of $z = 1.990\pm0.001$ while the redshift of Mrk~1456 is $z =
0.0476 \pm 0.0001$. The spectra, shown in Fig.~\ref{fig:specall},
cover the wavelength range from 3815\,\AA\ to 9205\,\AA, i.e. the
rest-wavelength range of the spectrum of Mrk~1456 is 3640\,\AA\ to
8785\,\AA\ and for the QSO 1276\,\AA\ to 3079\,\AA.

The measurements of the emission line fluxes were performed by fitting
Gaussians to the lines. The $1\sigma$ errors in the fluxes due to
photon noise were estimated using the error spectrum provided by SDSS
and assuming Poisson statistics. The $1\sigma$ error in the continuum
placement was taken to be one third of the mean noise of the continuum
in the spectral region around the measured lines. The total error for
each line measurement was determined by error propagation. For
H$\delta$, H$\gamma$, and H$\beta$, the line fitting included the
underlying stellar absorption. For the H$\alpha$ line, the absorption
was swamped by emission, therefore we adopted the correction method
recommended by \cite{hopkins2003}, for SDSS spectra, to calculate the
line flux. The $3\sigma$ upper limits for undetected lines were
calculated from the SDSS error spectrum.



We note that all spectroscopically derived parameters of the galaxy
pertain to the area surrounding its nucleus enclosed by the 3~arcsec
diameter fiber, which, at the distance of Mrk~1456, encompasses an
area of $\pi (2.8\,{\rm kpc}/2)^2=6.2\,{\rm kpc}^2$.

\begin{figure}
\includegraphics[angle=90,scale=.65]{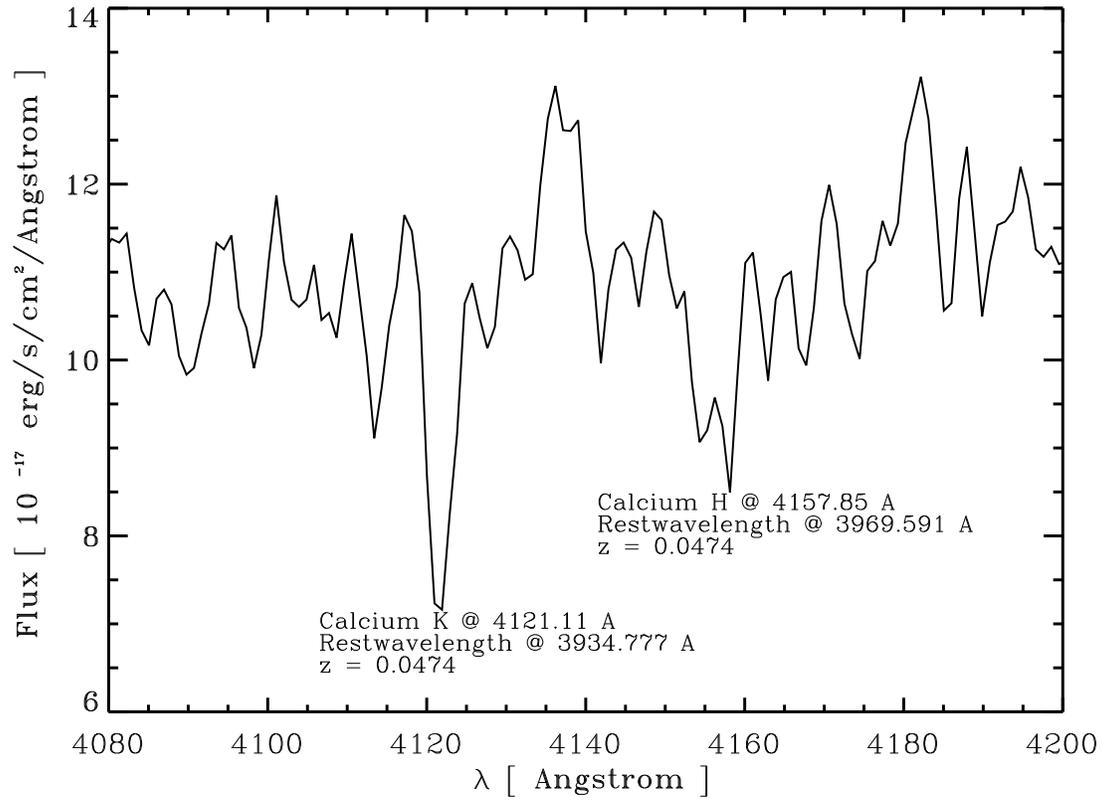}
\caption{Section of SDSS spectrum of the QSO from the SDSS
  homepage. The spectrum is smoothed with a boxcar of 3 pixel. We have
  labeled the calcium H \& K absorption lines which are observed at a
  redshift of z=0.0474.\label{fig:spec_qso}}
\end{figure}

Figure~\ref{fig:spec_qso} shows the calcium absorption lines caused by
the galaxy in the spectrum of the QSO. The equivalent widths (EW) and
width errors were measured by smoothing the spectrum with a boxcar of
3~pixels prior to the application of the line-fitting procedure
outlined above.





\section{Galaxy morphology, luminosity and size}

Mrk~1456 has an angular r-band Petrosian radius of 7.7$\pm$0.2\,arcsec
(or a physical radius of 7.0$\pm$0.2\,kpc). We used the code of
\cite{blanton2003} to k-correct the ugriz observations. They were
subsequently corrected for foreground and internal
absorption. Foreground reddening, $E(B-V) = 0.024$\,mag, and
absorption, $A_B =0.105$\,mag, were determined from the maps of
\cite{schlegel1998}. For the internal absorption correction, we
assumed the absorption derived for the nuclear region using the Balmer
lines (see next section) applies to the entire galaxy. Using the color
transformations of \cite{smith2002}, we find $M_B=-21.0\pm0.2$. This
makes Mrk~1456 a giant, rather than a dwarf (M$_B\gtrapprox-18$)
galaxy \citep{binggeli1985}. We derive $L/L^* = 1.1\pm0.2$ when
using $M_B^* =-20.9$ \citep{marinoni1999}.

Upon visual inspection of the images and spectrum of Mrk~1456, we find
that it resembles a spiral galaxy. By employing the criteria
established in \cite{kewley2001} to our measurements of the diagnostic
line ratios of [OIII]/H$\beta$, [NII]/H$\alpha$, [SII]/H$\alpha$, and
[OI]/H$\alpha$, we conclude that Mrk~1456 is a SFG, rather than an
AGN. The comparison of the spectrum with the templates from
\cite{kennicutt1992} suggests that it is a spiral of Sc subtype. The
inverse concentration index in the r~band (0.45), together with the
extinction-corrected Petrosian ugriz colors, place Mrk~1456 among the
Sc galaxies \citep{strateva2001, shimasaku2001}. Mrk~1456 was observed
in the Two Micron All Sky Survey (2MASS). Following \cite{Girardi},
its B-K color is consistent with the Sc subtype.

The following data allow us to address the inclination of the disk of
Mrk~1456.  The minor-to-major axis ratio in the r band resulting from
an exponential fit to the profile is 0.724$\pm$0.007. In the K band,
this axis ratio is 0.720. Both, r and K, measure the light from the
underlying older stellar population in the disk of Mrk~1456. Assuming
an axis ratio of 0.72 and following \cite{tully1977}, we derive an
inclination of 45$^o$ for Mrk~1456.

\section{Reddening and starformation rate}


The reddening of QSO spectra caused by intervening galaxies was
investigated by \cite{ostman2006}. They noted that Mrk~1456 and its
QSO as one of two very low-$z$ cases in their sample. They determine
that $R_V$ is close to the average Milky Way value of $R_V=3.1$.

The flux ratio $F(H\alpha)/F(H\beta) = 4.33 \pm 0.36$ (see
Tab.~\ref{tab_sdss_g}) is larger than the case B recombination value
of 2.86 for the ``standard'' $n_e =100$\,cm$^{-3}$ and $T_e = 10^4$\,K
HII region density and temperature \citep{osterbrock1989}. From this
Balmer-line ratio and using the reddening law of \cite{ccm89} we
derive $E(B-V)=0.42\pm0.08$\,mag, $A_V=1.3$\,mag, and $A_B=1.7$\,mag
assuming $R_V=3.1$.

Using \cite{kennicutt1998}, we calculate a nuclear star-formation rate
(SFR) of $0.68\pm0.13$\,M$_\odot$/yr from the H$\alpha$ luminosity
(corrected for stellar absorption and foreground and internal dust
absorption).

Following \cite{hopkins2003} we find the global SFR from
the Petrosian u-band magnitude, $1.3\pm0.5{\rm M}_\odot/{\rm yr}$.

%
%
%
%
\subsection{Oxygen abundance from spectral line indices}

We use the $O3N2$ index proposed by \cite{pettini2004}, $O3N2 = \log
\frac{{\rm[OIII]}\lambda 5007}{H\beta} / \frac{{\rm [NII]}\lambda
6583}{H\alpha}$, together with their empirical calibration to the
observed data (their Eq.~3), to calculate the oxygen abundance. We
derive an oxygen abundance of $12 + \log \frac{O}{H}
=8.6\pm0.1\pm0.14$ (where the first error denotes our measurement
error and the second error is the systematic error of the
calibration).

Using the solar abundance of \cite{holweger2001}, $12 + \log
\frac{O}{H} = 8.74\pm0.08$, we derive a solar-relative abundance of
${\rm [O/H]_{\rm II}}=-0.1\pm0.2$ \footnote{We will use the following
  notation: $[X/Y]=\log\frac{N(X)}{N(Y)} -
  \left.\log\frac{N(X)}{N(Y)}\right|_\odot$, $[X/Y]_{\rm I}$\ denotes
  abundance in the neutral gas phase and $[X/Y]_{\rm II}$\ denotes the
  abundance in the ionized gas phase.} for the nucleus of Mrk~1456.

This value is confirmed by using the $N2=\log \frac{{\rm [NII]}\lambda
6583}{H\alpha}$ index together with Eq.~2 of \cite{pettini2004}. We
derive $12 + \log \frac{O}{H} =8.5\pm0.2\pm0.18$ or ${\rm[O/H]_{\rm
II}}=-0.2\pm0.3$.

For further comparison, the R23 method \citep{mcgaugh1991, kobulnicky1999a},
which is calibrated on photoionization models, yields $12 + \log \frac{O}{H} =
8.7\pm0.2\pm0.10$ or a solar oxygen abundance.

In summary, all three methods indicate a solar metallicity for the
nucleus of Mrk~1456. This is in agreement with the observation of
\cite{shi2005} that the mean offsets between these three calibrators
are of the same order as their systematic errors.

\subsection{Electron Density and Temperature}
We use the IRAF/STSDAS package {\it nebular} \citep{shaw1995} with the
task {\it temden} to calculate the electron density ($n_e$) and
temperature ($T_e$) for the emission line spectrum. Some lines are so
weak that we can only measure an upper limit for the flux (see
Tab.~\ref{tab_sdss_g}). As a consequence, we can only determine 
the lower limit of the electron density
and the upper limit of the electron temperature. First we assume an
effective temperature of 10000\,K and calculate $n_e$ using the $\rm
[SII]6718,6732$ line ratio. Using that $n_e$, we derive an upper limit
for the temperature using the $\rm [SII]$ and $\rm [OIII]$
ratios. After two iterations we derive a lower limit for $n_e \ge
22$\,cm$^{-2}$ and an upper limit for the $T_e \le 7500$\,K.

We can also apply a reverse approach by assuming that the oxygen
abundance derived by the strong line indices is the true oxygen
abundance. Adding Eq.~3 and Eq.~5 of \cite{izotov2006} for the
abundances of the two ionization states O$^+$ and O$^{2+}$ and
assuming $\frac{O}{H}=\frac{O^+}{H^+}+\frac{O^{2+}}{H^+}$, we derived
a temperature of $T_e(O^{2+})=6800\pm1000$\,K, when using
$12+\log(O/H) =8.6\pm0.2$\ and an electron density of
$n_e(S^+)=22\pm2$\,cm$^{-2}$. This is consistent with the upper
limit derived above. In the following we will use this electron
density and temperature.

%
%
%
%
%
%
%
%
\subsection{Sulfur and Nitrogen Abundances}

Following \cite{izotov2006}, we calculate the abundances of $S^+$,
$S^{2+}$, and $N^+$\ and the corresponding ionization
correction factors (ICFs) from the emission line measurements given in
Tab.~\ref{tab_sdss_g}.
%
%
For sulfur we derive $12+\log\frac{\rm S^+}{\rm H^+}= 6.9\pm0.2$\ and
$12+\log\frac{\rm S^{2+}}{\rm H^+}= 7.4\pm0.4$. The
$ICF(S^++S^{2+})=0.95$. This leads to $12+ \log\frac{\rm S}{\rm H^+} =
7.5\pm0.5$ or $[S/H]_{\rm II}=0.3\pm0.3$ using the solar sulfur
abundance of $7.20\pm0.06$\ \citep{gs1998}.

For nitrogen we calculate $12+\log\frac{\rm N^+}{\rm H^+}= 7.8\pm0.3$\
and an $ICF(N^+)=1.60$. The total nitrogen abundance
$12+\log\frac{N}{H^+}=8.0$. With a solar abundance of $7.931\pm0.111$\
\citep{holweger2001} we derive ${\rm [N/H]_{II}}=0.1\pm0.3$. The
nitrogen over oxygen abundance is $\log N/O = -0.6$.

%
%
%
%
%
\section{Abundances on the line of sight to the QSO}

The QSO SDSS~J114719.90+522923.2 intercepts the disk of Mrk~1456 at a
distance that is smaller than its optical radius. We searched for
absorption lines of Ca H\&K and Na\,ID in the spectrum (3\,\AA\
resolution) and detected the Ca~H\&K lines (see
Fig.~\ref{fig:spec_qso}). The redshifts for Ca~II K and for Ca~II H are
$0.0474\pm0.0002$\ for each line. This compares well with the
emission-line derived redshift of $0.0476\pm0.0001$ for Mrk~1456 from
the SDSS database.

The measured rest equivalent widths are $1.24\pm0.15$\,\AA\ and
$0.76\pm0.20$\,\AA, respectively. Using the curve of growth method and
the oscillator strength of \cite{morton2003} we derive $\log N({\rm CaII})
= 13.2\pm0.3$\ using both lines.

\section{Discussion}

The characteristics of Mrk~1456 are interesting as it is both a
star-forming galaxy and a quasar absorption-line system. Here, we
discuss these characteristics and compare them to SBS~1543+593, the
only other DLA for which detailed studies of its emission and
absorption properties have been carried out.

\subsection{Mrk 1456 as a star-forming galaxy}

Mrk~1456 is a SFG with the morphology of a late spiral. The SFR in the
nucleus of Mrk~1456 is $0.1$~M$_\odot$~yr$^{-1}$~kpc$^{-2}$, while its
global SFR is $0.008$~M$_\odot$~yr$^{-1}$~kpc$^{-2}$. The young
stellar populations clearly dominate in the nucleus of
Mrk~1456. Indeed, according to the definition of \cite{balzano1983},
we may consider Mrk~1456 a starburst nuclear galaxy. Mrk~1456 has also
been classed a starburst nucleus by \cite{stepanian2005}.

Starburst nuclear galaxies show ionized gas in their inner regions
which is characterized by low excitation, \cite{contini2002}, and a
resulting high metallicity, \cite{coziole}. A comparison with the data
of \cite{contini2002} (shifted to our cosmology) shows that the
nuclear O/H and N/O ratios of Mrk~1456 are typical of a starburst
galaxy nucleus of its luminosity. Its N/O ratio is typical of its O/H
ratio \citep{shi2005}. Since Mrk~1456 is an SDSS galaxy, it is
expected to lie on the \cite{tremonti2004} metallicity-luminosity
relation. Figure~4 of Tremonti et al. illustrates the
metallicity-luminosity distribution of SDSS galaxies and the
differences in metallicity-luminosity calibrations that originate from
choosing other galaxy samples. Mrk~1456 has a low metallicity for its
luminosity, but lies, of course, within the SDSS galaxy
distribution. Mrk~1456 appears to lie closer to the \cite{contini2002}
metallicity-luminosity relation for UV-selected and HII galaxies
rather than on the SDSS relation. A possible interpretation is that
Mrk~1456 is shifted to a brighter magnitude for its metallicity
compared to normal spirals owing to the low M/L ratio of its newly
formed stellar populations \citep{contini2002}.

Our abundance measurements reflect the nuclear abundances only; we do
not have data in hand that would allow us to infer global abundances.
Because spiral galaxies are known to exhibit radial abundance
gradients with a negative slope that is particularly steep in late
subtypes, \cite{vilac}, our values are expected to overestimate the
global abundances of Mrk~1456, \cite{kewley2005}. The exact gradient
applicable to the case of Mrk~1456 can only be determined with future,
long-slit spectroscopy.

\subsection{Mrk 1456 as a QSO absorber and candidate DLA}

Studies of sightlines to QSOs enable us to probe the ISM of
intervening galaxies. This technique has been exploited to help us
better understand the ISM within our Galaxy
\citep[e.g.,][]{lauroesch1996, savage2000}.

We detected CaII absorption in the spectrum of the QSO SDSS
J114719.90+522923.2 at the wavelengths at which we expect them for
Mrk~1456. This establishes Mrk~1456 as a CaII absorber.

Two properties strongly suggest that Mrk 1456 is also a DLA galaxy,
namely, its strong CaII~K line and the small distance of the QSO from
the center of the galaxy.

\cite{wild2006} have investigated calcium absorption line systems on
the sightline of QSOs. They conclude that systems with $EW({\rm Ca\
II~K}~\lambda 3935) \gtrsim 0.68$\,\AA\ are a subset of the
DLA population. For Mrk~1456, we measure a Ca~K line $EW$ of
$1.24\pm0.15$\,\AA. Thus, Mrk~1456 is highly likely to be DLA galaxy
on accounts of the strength of the Ca absorption. Wild et al. surmise
that the regions giving rise to strong CaII absorption lines in
galaxies are small, 7-8~h$^{-1}$\,kpc. The impact parameter of
SDSS~J114719.90+522923.2 is 4.9\,kpc, consistent with this inference.

In their recent review, \cite{wolfe2005} summarized the HI properties
of spiral galaxies. The column densities decrease from $N(HI) \sim
10^{21}$\,cm$^{-2}$\ in the center to $N(HI) =1.8 \times
10^{20}$\,cm$^{-2}$\ at about $(1.5\pm0.5) R_H$ where $R_H$\ is the
Holmberg radius defined as the semi-major axis of the $26.5 \mu_B$
isophote in the B-band. We used the SDSS measured values for the
semi-major axis of the $25 \mu_g$ isophote and the gradient along the
major axis in the g-band in order to estimate the Holmberg radius for
Mrk~1546. We find that the Holmberg radius is approximately
7.3\,arcsec. Since the sightline to the QSO is only 5.3\,arcsec from
the center of the galaxy, the column density should exceed the
defining DLA column density, and we may expect that Mrk~1456 is a DLA
galaxy.

We also used the SFR calculated from the H$\alpha$
luminosity to estimate the nuclear HI column density of Mrk~1456 using
Eq.~6 from \cite{lanzetta2002} (which is the Schmidt law from
\citealt{kennicutt1998} but without the dust correction). Accordingly,
the central column density is estimated to be $N(HI)=1.8 \times
10^{22}$\,cm$^{-2}$.
 
It is interesting to note that \cite{rao2003} studied 14 DLA galaxies
with redshifts between 0.05 and 1, and found that low-luminosity
galaxies at small impact parameters dominate their sample. Being
an $L^*$ galaxy, Mrk~1456 adds a possible data point at the
high-luminosity end of the luminosity distribution of DLA galaxies.
Rao et al. also draw attention to the fact that the highest column
densities are observed in the galaxies with the lowest luminosities,
which they attribute to a possible selection bias, such that bright
galaxies with high column densities at small impact parameters are
missing from the sample. They suggest that this could happen because
the background QSO is dimmed too much by reddening when seen through
the disk of a luminous galaxy.

An alternate explanation for the existence of a bias against luminous
DLA galaxies may instead come about due to the fact that sightlines to
QSOs that pass through disks of nearby and thus, extended foreground
galaxies, have historically been missed because a standard cut that
has been applied in the assemblage of pre-SDSS quasar catalogs has
been that the object must appear stellar.

Our predication that Mrk~1456 is a DLA galaxy can be tested with
ultraviolet spectroscopy of the QSO at the wavelength of the Lyman
$\alpha$ line, or with 21-cm emission mapping of the galaxy.

\subsection{Comparison of Mrk~1456 and SBS~1543+593}

\begin{figure}[h]
\includegraphics[scale=.7]{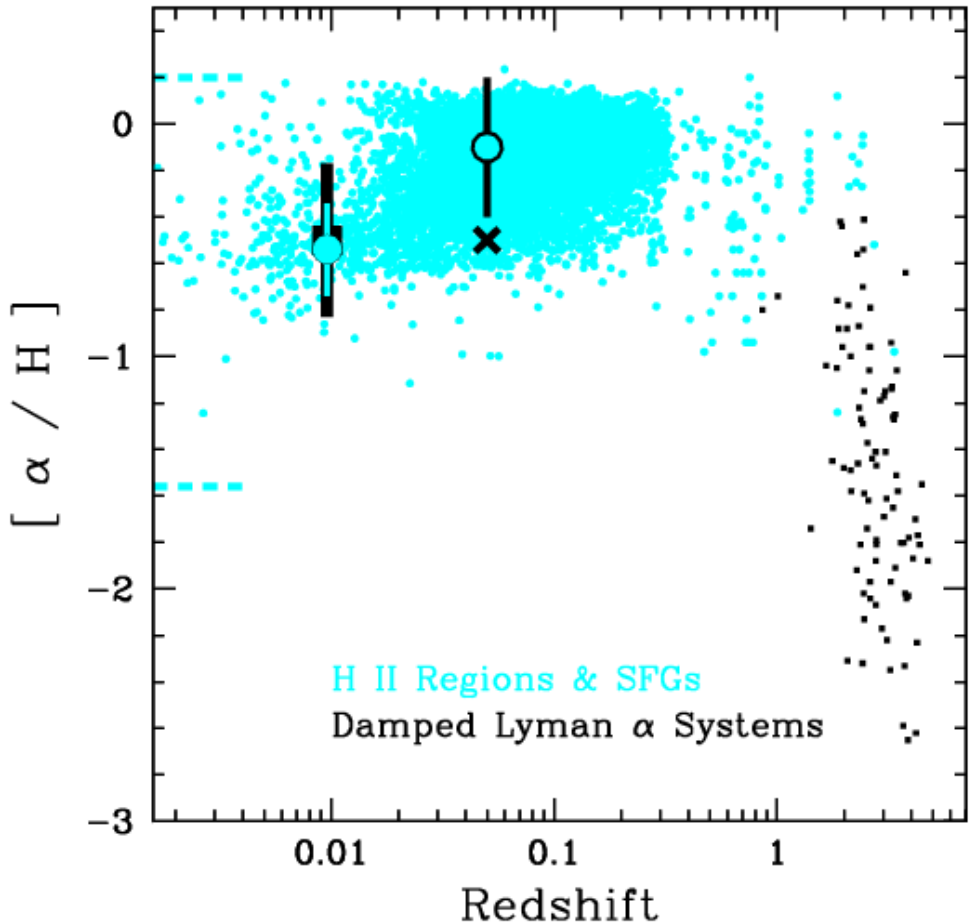}
\caption{Metallicity as a function of redshift. The dot and cross at
  redshift 0.05 are the metallicity measured in the center of Mrk~1456
  and the prediction to the sightline of the QSO if there is a metallicity 
  gradient (see text for details). The big black square
  symbol and the dot at redshift 0.01 are the metallicity measured in
  the neutral gas phase and in the HII region \#5 of SBS~1543+593 from
  \cite{sl2005}.  Small dark squares illustrate the abundances of DLAs and small
  bright dots those of HII-regions and SFGs (see text for details).}
\label{fig:oxi}
\end{figure}

The chemical enrichment of the ISM of galaxies is a critical indicator
of galaxy evolution. In Fig.~\ref{fig:oxi} we show with black squares,
the [O/H]$_{\rm I}$, [S/H]$_{\rm I}$, and [Si/H]$_{\rm I}$ ratios from
the compilation of DLA metallicities by \cite{prochaska2003d}, which
represent our knowledge of $\alpha$/H ratios in the neutral gas at
high redshift.

In order to connect the metallicities of DLAs with those of luminous
galaxies, we show with bright dots, [O/H]$_{\rm II}$ ratios for
high-redshift emission-line galaxies. The data are from
\cite{hoyos2005}, \cite{shapley2004}, \cite{steidel2004},
\cite{koo1994}, \cite{kobulnicky1999b}, \cite{lb2003},
\cite{rigo2000}, \cite{maier2004}, \cite{maier2005},
\cite{contini2002}, \cite{cardiel2003}, \cite{teplitz2000}, and
\cite{vm2004}, and cover the redshift range from about 0.3 to 3. If
these authors used the $T_e$ method or the \cite {PP2004} method to
calculate O/H ratios, we used their values. Otherwise, to be
consistent, we recalculated the O/H ratios from the original data
using the \cite {PP2004} method, because different metallicity
calibrations may yield different abundances \citep[e.g.,][]{shi2005}.

In order to extend the O/H ratios to low redshifts, we use the
metallicities for SDSS HII-regions and SFGs galaxies from
\cite{sl2005}, also calculated using the Pettini \& Pagel method. In
the local Universe, oxygen abundances have been determined using the
direct or $T_e$ method. I~Zw~18 is the local galaxy with the lowest
oxygen abundance on record, $-1.56\pm0.01$ \citep{IT1999}. The local H
II region with the highest known oxygen abundance, $0.2\pm0.2$,
resides in NGC 1232 \citep{Cast2002}. These boundaries on the
$z\approx 0$ metallicities are indicated by dashed lines.

Overplotted with a large square at a redshift of about 0.01, is the
absorption-derived [S/H]$_{\rm I}$\ abundance of SBS~1543+593
\citep{sl2005}. On top of that is found the emission-derived
[O/H]$_{\rm II}$\ abundance calculated using \cite{PP2004}, indicated
by a bright circle. The two measurements agree.

The emission-line derived nuclear [O/H]$_{\rm II}$\ ratio of Mrk~1456
is shown as the large dot at a redshift of about 0.05. Because
Mrk~1456 is much more luminous than SBS~1543+593, its oxygen abundance
is higher than that of SBS~1543+594, and similar to that of other
luminous SDSS galaxies at redshift 0.05. We do not know the
absorption-derived abundance of Mrk~1456 because it has not been
measured with ultraviolet spectroscopy. However, we can make some
inferences based on its emission-line derived abundance.

We have interpreted SBS~1543+593 as the ``proof of concept", showing
that emission- and absorption-derived measurements in principle do
yield consistent abundances of $\alpha$\ elements in the ISM of
external galaxies \citep{sl2005}. Based on this result, we
predict that the $\alpha$\ element abundances for Mrk~1456 measured
from absorption-line spectroscopy, will give the same result as its
emission-line derived one. 

In what follows, we consider that the value may actually turn out to
be lower, because the emission-line derived O/H ratios of spiral
galaxies decrease with radius and because the QSO sightline intercepts
the galaxy at 70\% of its radius.

Local dwarf galaxies, in which multiple HII region abundances as a
function of radius have been measured, appear to exhibit flat
abundance gradients \citep{kobulnicky1997}. SBS~1435+593 is a dwarf
galaxy. The sightline to the QSO intercepts the disk of SBS~1543+593
about 2.4\,arcsec (or $0.5 h_{70}^{-1}$\,kpc) to the N of the center
of the galaxy. The HII region which was used to gauge the ionized
gas-phase abundance, on the other hand, is located 14.7\,arcsec (or
2.9\,kpc) to the S of the galaxy's center. Despite their different
galacticentric distances, the two abundances agree, as expected based
on our current knowledge of abundance gradients in dwarfs.

In contrast, local giant galaxies, in which multiple HII region
abundances as a function of radius have been measured, appear to
exhibit negative abundance gradients \citep[e.g.][and references
therein]{vilac, ferguson1998, kewley2005}. In the case of Mrk~1456,
the impact parameter of the QSO is 4.9\,kpc.  If we extrapolate the
nuclear oxygen abundance of Mrk~1546 using the metallicity gradient of
\cite{ferguson1998}, $-0.09$\,dex/kpc, we predict an oxygen abundance
of $[O/H]\approx-0.5$\ on the sightline of the QSO. This abundance is
indicated by an x on Fig.~\ref{fig:oxi}. Doing the same for nitrogen
over oxygen (average abundance gradient of $-0.05$\,dex/kpc) we
estimate $[N/O]\approx0.0$ on the QSO sightline.

Interestingly, the O/H ratio for the neutral ISM of the giant spiral
Mrk~1456, estimated under the assumption of a steep radial abundance
gradient, is very similar to the one we observe in the dwarf spiral
SBS~1543+594. This would seem to support the conclusions reached by
\cite{chen2005} that low-redshift DLA galaxies are drawn from the
typical field population and not from a separate population of
low-surface-brightness dwarfs, and that their low metallicities may
arise naturally as a combination of gas cross section selection, which
favors large radii, and metallicity gradients.

We note that neither the abundance gradients of dwarf galaxies, nor
those of spiral galaxies, have yet been probed with absorption-line
spectroscopy. Only future ultraviolet spectroscopy can tell us the
true absorption-line derived $\alpha$ element abundance on the
sightline to SDSS~J114719.90+522923.2.

\section{Conclusion}

Mrk~1456 is an $L^*$ spiral galaxy with a starburst nucleus. The
emission-line spectrum of the nucleus yields a SFR of
about 0.7\,M$_\odot$/yr. The oxygen abundance in the warm ISM of
Mrk~1456 is about solar, consistent with its luminosity. We predict
that the oxygen (or sulfur) abundance in the ISM on the sightline to
SDSS~J114719.90+522923.2 is either approximately solar, based on our
observations of the SBS~1543+593-HS~1543+5921 galaxy-QSO pair, or it
may be only about $30\%$\ solar if the galaxy exhibits a radial
abundance gradient similar to that of local spirals. Future
ultraviolet spectroscopy is needed to test these predictions.

The sightline to the background QSO SDSS~J114719.90+522923.2
intercepts the disk of Mrk~1456 within its Holmberg radius. Detection
of the CaII H\&K lines in the QSO's spectrum designates the galaxy a
CaII absorber. The small impact parameter of the QSO's sightline,
4.9\,kpc, as well as the strength of the CaII~K absorption establish
Mrk~1456 as an extremely likely candidate for a DLA galaxy. This
prediction can be tested with future ultraviolet spectroscopy or with
21-cm mapping. Once proven to be a DLA galaxy, Mrk~1456 will make an
interesting, high-luminosity addition to the very small sample of
low-redshift DLA galaxies \citep{chen2005, rao2003}.


\acknowledgments {\bf Acknowledgments}\\ We thank Alex
Fiedler, Andrew Hopkins, Simon Krughoff, Ryan Scranton and Ramin Scribba for the
helpful discussions. Funding for the Sloan Digital Sky Survey (SDSS)
has been provided by the Alfred P. Sloan Foundation, the Participating
Institutions, the National Aeronautics and Space Administration, the
National Science Foundation, the U.S. Department of Energy, the
Japanese Monbukagakusho, and the Max Planck Society. The SDSS Web site
is http://www.sdss.org/. The SDSS is managed by the Astrophysical
Research Consortium (ARC) for the Participating Institutions. The
Participating Institutions are The University of Chicago, Fermilab,
the Institute for Advanced Study, the Japan Participation Group, The
Johns Hopkins University, the Korean Scientist Group, Los Alamos
National Laboratory, the Max-Planck-Institute for Astronomy (MPIA),
the Max-Planck-Institute for Astrophysics (MPA), New Mexico State
University, University of Pittsburgh, University of Portsmouth,
Princeton University, the United States Naval Observatory, and the
University of Washington. This research has made use of the
NASA/IPAC Extragalactic Database (NED) which is operated by the Jet
Propulsion Laboratory, California Institute of Technology, under
contract with the National Aeronautics and Space Administration.
This publication makes use of data products from the Two Micron All
Sky Survey (2MASS), which is a joint project of the University of
Massachusetts and the Infrared Processing and Analysis
Center/California Institute of Technology, funded by the National
Aeronautics and Space Administration and the National Science
Foundation.

\clearpage


%
\clearpage
\clearpage
\clearpage

\clearpage

\clearpage

\begin{table}
\small
\begin{center}
\caption{Emission line fluxes of Mrk~1456 in the SDSS spectrum. Errors
  are $1 \sigma$ measurement errors and upper limits are at $3
  \sigma$. We calculated dereddened luminosities using the reddening
  law of \cite{ccm89}. \label{tab_sdss_g}}
\begin{tabular}{lcrrl}
\tableline\tableline
Species & $\lambda_0$    & measured & dereddened & Remark \\
        &                & Flux     & Luminosity &        \\      
        &  [\AA]         & $10^{-16}\left[{\rm \frac{erg}{cm^2 s}}\right]$ & $10^{40}\left[{\rm \frac{erg}{s}}\right]$&\\ 
\tableline
$\rm [OII]$  & 3728  & $35.25\pm0.2$  &  $11.0\pm4.0$ & \\ 
$\rm [SII]$  & 4068  & $<0.60$        &  $<0.17$ & \\
$\rm [SII]$  & 4076  & $<0.60$        &  $<0.17$ & \\
H$\delta$    & 4101  & $1.48\pm0.2$   &  $ 0.41\pm0.19$ & [1]\\
H$\gamma$    & 4341  & $4.00\pm0.2$   &  $ 0.99\pm0.36$ & [1]\\
$\rm [OIII]$ & 4363  & $<0.50$        &  $<0.12$ &  \\
H$\beta$     & 4862  & $15.15\pm0.2$  &  $ 3.02\pm 0.86$ & [1]\\
$\rm [OIII]$ & 4959  & $2.98\pm0.1$   &  $ 0.57\pm 0.19$ &  \\
$\rm [OIII]$ & 5007  & $9.33\pm0.1$   &  $ 1.8\pm 0.5$ &  \\
HeI          & 5877  & $1.60\pm0.1$   &  $ 0.24\pm 0.07$ & [2]\\
$\rm [OI]$   & 6300  & $2.25\pm0.1$   &  $ 0.31\pm 0.08$ &  \\
$\rm [SIII]$ & 6312  & $0.26\pm0.1$   &  $ 0.04\pm 0.02$ &  \\
$\rm [NII]$  & 6548  & $6.80\pm0.2$   &  $ 0.90\pm 0.20$ & [3]\\
H$\alpha$    & 6563  & $65.65\pm0.3$  &  $ 8.64\pm 1.69$ & [4]\\       
$\rm [NII]$  & 6584  & $19.40\pm0.3$  &  $ 2.55\pm 0.52$ &  \\
$\rm [SII]$  & 6718  & $13.30\pm0.2$  &  $ 1.70\pm 0.34$ & [5]\\
$\rm [SII]$  & 6732  & $9.47\pm0.2$   &  $ 1.2\pm 0.3$ & [5]\\
$\rm [OII]$  & 7320  & $<0.45$        &  $<0.05$ &  \\
$\rm [OII]$  & 7330  & $<0.45$        &  $<0.05$ &  \\
\tableline
\end{tabular}
\newline
[1]~abs.+em. deblended with 2 Gaussians, [2]~near abs. line, [3]~3
em. lines deblended,\newline
[4]~and H$\alpha$ abs. corrected, [5]~deblended with a Gaussian

\end{center}
\end{table}

\clearpage

\end{document}